# Low temperature rate constants for the $N(^4S) + CH(X^2\Pi_r)$ reaction. Implications for $N_2$ formation cycles in dense interstellar clouds.

**Julien Daranlot,[a] Xixi Hu,[b] Changjian Xie,[b] Jean-Christophe Loison,[a] Philippe Caubet,[a] Michel Costes,[a] Valentine Wakelam,[c] Daiqian Xie,[b],* Hua Guo[d],* and Kevin M. Hickson [a],***

Rate constants for the potentially important interstellar $N(^4S) + CH(X^2\Pi_r)$ reaction have been measured in a continuous supersonic flow reactor over the range 56 K ≤ T ≤ 296 K using the relative rate technique employing both the $N(^4S) + OH(X^2\Pi_i)$ and $N(^4S) + CN(X^2\Sigma^+)$ reactions as references. Excess concentrations of atomic nitrogen were produced by the microwave discharge method upstream of the Laval nozzle and CH and OH radicals were created by the *in-situ* pulsed laser photolysis of suitable precursor molecules. In parallel, quantum dynamics calculations of the title reaction have been performed based on accurate global potential energy surfaces for the $1^3A'$ and $1^3A''$ states of HCN and HNC, brought about through a hierarchical construction scheme. Both adiabatic potential energy surfaces are barrierless, each one having two deep potential wells suggesting that this reaction is dominated by a complex-forming mechanism. The experimental and theoretical work are in excellent agreement, predicting a positive temperature dependence of the rate constant, in contrast to earlier experimental work at low temperature. The effects of the new low temperature rate constants on interstellar $N_2$ formation are tested using a dense cloud model, yielding $N_2$ abundances 10-20 % lower than previously predicted.

## 1. Introduction

Reactions between two neutral radical species are involved in the chemistry of a wide variety of environments ranging from combustion processes at high temperature to complex molecule formation in the coldest, darkest regions of interstellar space. For a reaction to be considered important at temperatures characteristic of dense interstellar clouds which are often lower than 10 K, transient species formed along the path or paths leading exothermically to products must occur at energies lower than the reagent asymptote. Reactions presenting an activation barrier need only to be considered if they can occur by tunnelling through the barrier and endothermic processes are inhibited. Moreover, in these objects, the total density (gas + dust) is sufficiently high to prevent light from reaching the cloud core so that photodissociation processes initiated by external photons can be ignored. Much of the chemistry is initiated by cosmic rays, through ionisation of the hydrogen molecule whereupon $H_2^+$ reacts with neutral $H_2$ yielding the $H_3^+$ ion. $H_3^+$ itself plays a key role in dense cloud chemistry as it reacts with ground state atomic radicals present at large relative abundances ($10^{-4}$ – $10^{-5}$ with respect to total hydrogen ($nH + 2nH_2$)) to trigger a chain of reactions eventually leading to complex molecule formation. Both atomic carbon and oxygen react in this way to form neutral hydride species. The situation is quite different for ground state atomic nitrogen as reaction with $H_3^+$ is prohibited due to an activation barrier for one reaction pathway[1,2] whilst a second pathway is endothermic.[3] Instead atomic nitrogen reacts primarily with other small neutral radicals present in dense clouds

$N + OH \rightarrow H + NO$ (1) Mechanism (A)

$N + NO \rightarrow N_2 + O$ (2)

$N + CH \rightarrow CN + H$ (3a) Mechanism (B)

$N + C_2 \rightarrow CN + C$ (3b)

$N + C_2N \rightarrow CN + CN$ (3c)

$N + CN \rightarrow N_2 + C$ (4)

Until recently, many of these reactions were predicted to occur rapidly at 10 K (by the extrapolation of kinetic data recorded at higher temperatures), promoting the efficient conversion from N to $N_2$ at low temperatures. Astrochemical models predicted that molecular nitrogen should be the dominant reservoir of interstellar nitrogen with an abundance five orders of magnitude lower than that of total hydrogen. Although this hypothesis has been difficult to validate given that $N_2$ possesses no permanent dipole moment, indirect[4] and direct[5] observations indicate that real $N_2$ abundances could be substantially lower. Recent experimental and theoretical studies of reactions (1),[6] (2)[7-9] and (4)[10,11] support this idea by demonstrating that these three reactions could be slower than previously thought at 10 K. When these new rates are introduced into a dense cloud model[10] simulating both gas phase and surface reactions on interstellar grains, the derived gas-phase abundance of $N_2$ falls. Instead of reacting in the gas-phase, atomic nitrogen preferentially depletes onto interstellar grains where surface catalysed reactions with atomic hydrogen mostly lead to the formation of $NH_3$ ices as the main reservoir of interstellar nitrogen. Nevertheless, the efficiency of mechanisms leading to gas-phase CN formation, denoted by (3), remain largely unquantified at low temperatures. Whilst NO radicals are essentially produced only by reaction (1) in dense interstellar clouds, allowing us to quantitatively assess $N_2$ production by mechanism (A), CN radicals involved in mechanism (B) have several potential sources, one of the most important being reaction (3a). Current databases[12-14] use a large rate constant for this reaction equal to 2.3 × $10^{-10}$ $cm^3$ $molecule^{-1}$ $s^{-1}$, based on the extrapolation of measurements by Brownsword *et al.*[15] between 216 and 584 K. Current models tend to overestimate the CN abundance (models[16,10] predict abundances in the range $10^{-7}$ – $10^{-8}$ whereas observations lead to abundances in the range $10^{-9}$ – $10^{-10}$ [17,18]) at times which are assumed to best represent the age of a typical dense molecular cloud.

The present study reports the results of kinetics measurements of the N + CH reaction to temperatures as low as 56 K. The experiments are complemented by quantum scattering calculations based on newly developed accurate potential energy surfaces (PESs) for the two lowest lying triplet states of the HCN intermediate complex, allowing temperature dependent rate constants to be calculated. The experimental and theoretical methodologies used to investigate this important atom-radical reaction are outlined in section 2. Section 3 presents a comparison of the experimental and theoretical results with previous work whilst the conclusions and astrophysical implications are developed in section 4.

## 2. Methodology

### Experimental methods

Experiments were performed using a miniaturized continuous supersonic flow reactor based on the original design by Rowe *et al.*[19] The main features of the present system have been described previously.[6,10] Briefly, five Laval nozzles were used to perform kinetic measurements at specified temperatures of 56 K, 86 K, 146 K, 166 K and 167 K. Measurements were performed at 296 K by removing the nozzle and by significantly reducing the flow velocity in the reactor to eliminate pressure gradients in the observation region. Ground-state atomic nitrogen ($N^4S$) was generated upstream of the Laval nozzle by microwave discharge. A Vidal type microwave discharge cavity[20] operating at 2.45 GHz and up to 200 W was mounted on the sidearm of a Y shaped quartz inlet tube entering the Laval nozzle reservoir. Excess concentrations of $N(^4S)$ as high as 3.6 × $10^{14}$ $cm^{-3}$ could be produced in the cold flow as estimated from fits to the kinetic decays. An earlier investigation in our laboratory[6] showed that excited states of atomic nitrogen $N(^2P^0)$ and $N(^2D^0)$ were either quenched or removed by reaction before reaching the cold supersonic flow under similar conditions to the present experiments. A secondary consequence of the high discharge power used to produce excess atomic nitrogen was an elevated gas temperature within the nozzle reservoir. As the reservoir temperature was also found to vary as a function of the molecular nitrogen flow through the discharge, the discharge power was varied for different molecular nitrogen flows to maintain a constant reservoir temperature with a specified nozzle. The upstream temperature was measured by a type K thermocouple inserted into the reservoir in separate calibration experiments performed under identical conditions to the main experiments. At the same time, the supersonic flow characteristics were recorded through measurements of the impact pressure using a Pitot tube to verify that the microwave discharge did not cause perturbations. The measured reservoir temperatures, impact and stagnation pressures were subsequently used to calculate the temperature, density and velocity profiles of the supersonic flow.

Measurements of the rate constant for reaction (3a) were performed using the relative rate method, using the previously measured rates of the N + OH reaction as a reference.[6] In addition, a limited number of experiments were conducted at 296 K and at 146 K using the N + CN reaction as a reference by observing the temporal evolution of the CN product of the N + CH reaction. OH radicals in the $X^2\Pi_i$ state were generated by the *in-situ* pulsed photolysis of $H_2O_2$ with a 10 Hz frequency quadrupled Nd:YAG laser at 266 nm with ~ 27 mJ of pulse energy. $H_2O_2$ was introduced into the reactor by bubbling a small flow of carrier gas through a 50% weight mixture of $H_2O_2$ / $H_2O$. An upper limit of $6.7 \times 10^{11}$ molecule $cm^{-3}$ was estimated for the gas phase concentration of $H_2O_2$ in the supersonic flows from its saturated vapour pressure, providing OH concentrations lower than $8.0 \times 10^9$ molecule $cm^{-3}$ from calculations of the photodissociation efficiency of $H_2O_2$. Simultaneously, CH radicals in the $X^2\Pi_r$ state were generated by the 266 nm multiphoton dissociation of $CHBr_3$ molecules entrained in the flow by bubbling a small flow of carrier gas through a sample of liquid $CHBr_3$ held in a separate container. $CHBr_3$ has already been demonstrated as a reliable source of CH radicals by earlier kinetic studies.[21,22] An upper limit of $1.3 \times 10^{12}$ molecule $cm^{-3}$ was estimated for the gas phase concentration of $CHBr_3$ in the supersonic flows from its saturated vapour pressure. It should be noted that the gas-phase concentrations of precursor molecules $CHBr_3$ and $H_2O_2$ are substantially lower than those used in our other recent experiments. The use of small precursor densities was necessary during these experiments for two separate reasons. Firstly, the CH radical is highly reactive[21,22] and was observed to react rapidly with both $H_2O_2$ (and / or $H_2O$) and $CHBr_3$ in the cold supersonic flow. Secondly, OH radicals were seen to be produced by one (or several) secondary reactions as observed from non-zero baseline values of the OH signal at long times in the presence of higher concentrations of $CHBr_3$, $H_2O_2$ (and / or $H_2O$) and atomic nitrogen; both precursor molecules needed to be present to observe the secondary OH signal. To identify the source of secondary OH, test experiments were performed during which the mixture $H_2O_2$ / $H_2O$ was replaced by pure $H_2O$. These tests revealed that secondary OH radicals were still produced indicating that $H_2O$ was undoubtedly involved in OH production from secondary reactions (although this does not rule out the possibility that $H_2O_2$ also leads to secondary OH production). Whilst the reaction of CH with $H_2O$ occurs rapidly at room temperature ($k_{CH+H2O}$(298 K) ≈ 2-3 × $10^{-11}$ $cm^3$ $molecule^{-1}$ $s^{-1}$) and above[23,24] and recent statistical calculations[25] predict that the rate should increase below room temperature, this reaction is not thought to result in the formation of OH radicals by hydrogen abstraction as this pathway is endothermic. One potential source of OH radicals in the first excited state ($A^2\Sigma^+$) could be the CH + $O_2$, reaction ($O_2$ coming from the thermal decomposition of $H_2O_2$), however this cannot explain OH production in the presence of pure $H_2O$. Despite our best efforts, the secondary OH source (or sources, necessarily involving both precursor molecules) could not be identified. Nevertheless, the initial $CHBr_3$ and $H_2O_2$ / $H_2O$ concentrations were reduced to a minimum so that the OH signal always decayed to zero at long times. Under these conditions the loss of CH through reaction with $CHBr_3$, $H_2O_2$ and $H_2O$ was always much smaller (and constant) than the corresponding loss with N($^4S$). Moreover, supplementary experiments which employed the N + CN reaction as the reference avoided these complications given that $H_2O_2$ and $H_2O$ were absent from the system. Given the estimated minor reagent densities, reactions between CH and OH and between CH and CN had a negligible effect on the kinetic analysis.

The probe laser system for the detection of both OH and CN radicals by laser-induced fluorescence (LIF) have been described previously[26,10] with radical fluorescence being monitored using using a UV sensitive photomultiplier tube (PMT) coupled with the appropriate narrowband interference filter and a boxcar integration system. CH radicals were also followed by LIF via the R1-branch lines of the (0, 0) $A^2\Delta \leftarrow X^2\Pi_r$ band. For this purpose, two different pulsed lasers were used. In earlier experiments, the third harmonic 355 nm radiation of a pulsed single longitudinal mode Nd:YAG laser was used to pump an OPO system to produce tuneable radiation around 430.5 nm. In later experiments, 430.5 nm radiation was generated by frequency doubling the 861 nm output of a Nd:YAG pumped dye laser in a type I beta barium borate crystal. CH radicals were observed via the $A^2\Delta \rightarrow X^2\Pi_r$ (0, 1) band using a second UV sensitive PMT coupled with a 3 nm FWHM interference filter centered on 488 nm and a second boxcar integrator. Both probe lasers were coaligned and counterpropagated with respect to the photolysis laser along the axis of the supersonic flow.

LIF signals from the reactant OH (or product CN) and CH radicals were recorded simultaneously. For a given atomic nitrogen concentration, 30 datapoints were accumulated at each time interval with temporal profiles consisting of a minimum of 49 time intervals. This procedure was repeated several times at each atomic nitrogen concentration producing a minimum of 51 individual decay profiles for each temperature. As CH and OH signals were potentially non-zero in the absence of the photolysis laser (due to the possible upstream production of CH and OH radicals by the microwave discharge) several time points were recorded by firing the probe lasers before the photolysis laser. The LIF intensities were measured at a fixed distance from the Laval nozzle; the chosen distance corresponded to the maximum displacement from the nozzle for optimal flow conditions to be still valid thus allowing us to exploit the fluorescence signals over as large a period as possible.

All gases were flowed directly from cylinders with no further purification prior to usage. The carrier gas and precursor flows were all passed into the reservoir *via* digital mass flow controllers. The controllers were calibrated using the pressure rise at constant volume method for the specific gas used.

**Theoretical methods**

As shown in Figure 1, the reaction between N($^4$S) and CH($X^2\Pi_r$) leads to H($^2$S) + CN($X^2\Sigma^+$) and H($^2$S) + CN($A^2\Pi$) products on two separate triplet state PESs. To understand the reaction dynamics and kinetics, the $1^3A'$ and $1^3A''$ PESs have been mapped out at the Davidson corrected multi-reference configuration interaction (MRCI+Q) level with the state-averaged completed active space self-consistent field (CASSCF) wavefunctions.[27-29] The CASSCF calculations were performed with three states ($1^3A'$, $2^3A'$ and $1^3A''$) and a full-valence active space with 10 electrons distributed in 9 orbitals with the 1s orbitals of the carbon and nitrogen kept doubly occupied but fully optimized. All *ab initio* calculations were carried out with the MOLPRO 2010 package.[30]

The PESs for $1^3A'$ and $1^3A''$ electronic states were spline fitted in the internal coordinates using the hierarchical construction scheme suggested by Fu *et al.*[31] About 37000 *ab initio* points using the correlation-consistent polarized valence triple zeta (AVTZ) basis set were calculated to generate a smooth PES for each state. Higher accuracy PESs were obtained by adding the energy differences between the AV5Z and AVTZ PESs with 4240 points. These PESs indicate that both reaction pathways have no intrinsic barriers and are dominated by potential wells corresponding to the HCN and HNC isomers, which suggest a complex-forming reaction mechanism.[32]

The reaction probabilities for the two reaction channels were obtained using the Chebyshev real wave packet method.[33-35] Partial waves up to $J$=80 for the $1^3A'$ state and $J$=95 for the $1^3A''$ state were included to converge the integral cross sections (ICSs) shown in Figure 2 to 0.20 eV and the rate constants to 600 K. Finally, the rate constant for reactants in a specified initial state is given by a Boltzmann weighted average of the ICSs as a function of the collision energy. The final total rate constant to be compared with the experimental one is simply the sum of the rate constants for the two individual reaction channels. It is important to note that statistical models[36] that assume complete energy randomization almost certainly fail for this small molecular system due to the significant non-reactive scattering flux. Full quantum dynamics calculations are required to obtain accurate rate constants.[32] A more detailed description of the *ab initio* calculations, potential fitting and dynamical calculations can be found in the Supporting Information (SI) file.

## 3. Results and Discussion

Examples of the variation of the OH and CH radical fluorescence signal as a function of time recorded simultaneously at 296 K are given in Figure 3A. Atomic nitrogen was held in large excess with respect to the OH and CH radical concentrations for experiments conducted using reaction (1) as the reference, so that simple exponential fits to the OH and CH temporal profiles yielded the pseudo-first order rate constants for the N + OH and N + CH reactions, $k_{N+OH}'$ and $k_{N+CH}'$ respectively. This procedure was repeated at several different atomic nitrogen concentrations by varying the $N_2$ flow through the microwave discharge and / or by changing the microwave power. The values of $k_{N+CH}'$ obtained in this way for a range of atomic nitrogen concentrations were plotted as a function of the corresponding values of $k_{N+OH}'$ yielding a straight line for all experiments performed above 100 K. However, the ratio of the first-order rate constants for the target and reference reactions was seen to deviate from linearity at the lowest temperatures and in the presence of high [$N_2$] as a result of the CH + $N_2$ + M termolecular association reaction, which is known to become much more efficient below 100 K.[37] As described above, the standard experimental procedure usually consists of varying the $N_2$ flow through the microwave discharge to produce a range of atomic nitrogen concentrations. However this procedure was modified for experiments conducted below 100 K to maintain a constant $N_2$ concentration. As the $N_2$ flow through the microwave cavity was reduced, the "missing" $N_2$ was added through the main inlet instead, through a careful cross calibration of the two mass flow controllers involved. As the total $N_2$ flow was held constant, a linear relationship between the rate constants for the target and reference reactions was maintained, with the first order loss rate of CH with $N_2$ contributing to the value of the y-axis intercept. Two plots of the rate constants obtained using this procedure at 56 K are presented in Figure 4. The plots have different y-axis intercept values due to a twofold reduction in the precursor molecule concentrations between the two experiments in order to check for competing secondary reactions. For experiments performed at 80 K, the effect of the CH + $N_2$ reaction on kinetic decays was so small that experiments employing both constant and variable [$N_2$] were used in the final analysis. Weighted linear least squares fits to the data yielded the ratio of the two rate constants $k_{N+CH}'$ / $k_{N+OH}'$ at a given temperature from the slopes.

Examples of the temporal evolution of the reactant CH and product CN radical fluorescence signals recorded at 146 K are presented in Figure 3B. The interpretation of experiments monitoring the CN product of reaction (3a) was more complicated for two main reasons. Firstly, as CN is not initially present in the reactor but is instead produced by reaction (3a) before being consumed by reaction (4), the CN radical fluorescence intensity as a function of time can be described by the following formula considering consecutive pseudo-first order reactions

$$[\mathrm{CN}] = \frac{k_{N+CH}{}' [\mathrm{CH}]_0}{k_{N+CN}{}' - k_{N+CH}{}'} \left( e^{-k_{N+CH}{}' t} - e^{-k_{N+CN}{}' t} \right)$$

Secondly, the reaction of N($^4$S) with CH($^2\Pi_r$) occurs over two triplet surfaces,[38] leading exothermically in the absence of a barrier to the formation of ground state H($^2$S) atoms and CN radicals in both the ground $X^2\Sigma^+$ and the first excited $A^2\Pi$ states.[39] CN($A^2\Pi$) radicals formed by reaction (3a) will then either react with excess atomic nitrogen at a rate which might be comparable to the N + CN($X^2\Sigma^+$) one given that the spin multiplicity is the same for both reactions, or they will radiate back to the ground $X^2\Sigma^+$ state. As the lifetime of CN($A^2\Pi$) is relatively long (~ 12 μs)[40] this radiative process interferes with the simple consecutive kinetic analysis presented above. Numerical simulations of all the important reactive and radiative processes involved indicate that in the presence of high [N], only the fast rising part of the CN($X^2\Sigma^+$) fluorescence signal should be affected, shifting the peak CN signal to longer times whereas at low [N], radiative relaxation from the $A^2\Pi$ state contributes to the CN($X^2\Sigma^+$) fluorescence signal at a relatively low level throughout the entire temporal profile. In our recent experiments measuring the relative rates of the N + OH and N + NO reactions,[6] pseudo-first order rate constants for the N + NO reaction, $k_{N+NO}'$, were extracted by fixing the pseudo-first order rate constants for the N + OH reaction, $k_{N+OH}'$, describing the rising part of the NO fluorescence signal to the values determined from fitting to OH temporal profiles obtained simultaneously. Unfortunately, a similar analysis could not be performed in the present case as the equivalent rising part of the CN signal was substantially slower than expected from $k_{N+CH}'$ values recorded simultaneously (except in the presence of low [N]). Nevertheless, it was possible to extract the pseudo-first order rates for the N + CN reaction, $k_{N+CN}'$, albeit with a greater uncertainty by allowing both fitting parameters to vary in the analysis. Given the potential for significant error in fitting to such profiles, the results of experiments measuring N + CH rates relative to N + CN ones were not used to derive the overall temperature dependence of the N + CH reaction. Temperature dependent rate constants for the N + CH reaction were obtained by multiplying the ratios obtained from the slopes of plots similar to the ones presented in Figure 4 by the values of the rate constants previously obtained for the reference reaction at the corresponding temperature ($k_1$(56-300 K) = $4.5 \times 10^{-11}$ cm$^3$ molecule$^{-1}$ s$^{-1}$ and $k_4 = (8.8 \times 10^{-11}) \times (T/300)^{0.42}$ cm$^3$ molecule$^{-1}$ s$^{-1}$). These values are given in Table 1 alongside other relevant information and are shown as a function of temperature in Figure 5.

The calculated rate constants for the reaction of $N(^4S)$ with $CH(X^2\Pi_r)$ in the specific initial state ($v_i$=0, $j_i$=0) are also presented in Figure 5. Higher vibrational states of CH were not considered, as less than 0.2 % of total population is found in excited states below 600 K. Moreover, we assume that the rate constant is relatively insensitive to the CH rotational state given the complex forming nature of the reaction. It should be noted that the total rate constant has contributions from both the $1^3A'$ and $1^3A''$ pathways, with the latter being the major contributor above 150 K. Below 150 K, the relative contribution of the $1^3A''$ pathway falls as the electronic statistical weighting factor tends to zero at the lowest temperature, predicting a value of $k_{3a}$(10 K) = 0.8 × $10^{-11}$ $cm^3$ $molecule^{-1}$ $s^{-1}$.

A simple A × $(T/300)^B$ fit to the present experimental data and the data of Brownsword *et al.*[15] at room temperature and above yields a temperature dependence B = (0.41 ± 0.05) with A = (1.4 ± 0.4) × $10^{-10}$ $cm^3$ $molecule^{-1}$ $s^{-1}$. As the rate constant varies weakly with temperature, extrapolation allows us to estimate a value of $k_{3a}$(10 K) = 3.5 × $10^{-11}$ $cm^3$ $molecule^{-1}$ $s^{-1}$. Taking into consideration both the experimental and theoretical predictions leads us to recommend a value for $k_{3a}$(10 K) = (2.5 ± 2.0) × $10^{-11}$ $cm^3$ $molecule^{-1}$ $s^{-1}$, almost an order of magnitude smaller than the value of 2.3 × $10^{-10}$ $cm^3$ $molecule^{-1}$ $s^{-1}$ recommended by current astrochemical databases.[12-14]

At 296 K, $k_{3a}$ = (1.38 ± 0.15) × $10^{-10}$ and (1.21 ± 0.17) × $10^{-10}$ $cm^3$ $molecule^{-1}$ $s^{-1}$ from measurements employing reaction (1) and reaction (4) as the reference respectively, in good agreement with the value given by the present calculations of 1.22 × $10^{-10}$ $cm^3$ $molecule^{-1}$ $s^{-1}$ and earlier measurements by Brownsword *et al.*[15] at 294 K of (1.55 ± 0.14) × $10^{-10}$ $cm^3$ $molecule^{-1}$ $s^{-1}$. The only other previous measurement of reaction (3a) by Messing *et al.*[41] yielded a rate constant at 298 K of (2.1 ± 0.5) × $10^{-11}$ $cm^3$ $molecule^{-1}$ $s^{-1}$, almost an order of magnitude smaller than the present work. The various reasons for such a large discrepancy and the relative virtues and drawbacks of the work of Messing *et al.*[41] are described in some detail by Brownsword *et al.*[15] so no further explanation is required here. Below room temperature, the experimental rate constants obtained in this study are in excellent agreement with the theoretical ones predicting a significant fall off in the rate constant below 100 K. In contrast, the present experimental and theoretical results disagree with the only other experimental measurement at low temperature by Brownsword *et al.*[15] These authors measured a rate constant for reaction (3a) at 216 K $k_{3a}$(216 K) = (1.77 ± 0.26) × $10^{-10}$ $cm^3$ $molecule^{-1}$ $s^{-1}$ which compares with a value of (1.22 ± 0.1) × $10^{-10}$ $cm^3$ $molecule^{-1}$ $s^{-1}$ taken from the fit to our experimental data at 216 K. Although this difference is only just outside of the combined experimental error bars, the fitting parameters derived by Brownsword *et al.*[15] predict a negative temperature dependence of the rate constant with values of $k_{3a}$(56 K) and $k_{3a}$(10 K) equal to 1.9 and 2.3 × $10^{-10}$ $cm^3$ $molecule^{-1}$ $s^{-1}$ respectively, in stark contrast to our measured value of $k_{3a}$(56 K) = (5.95 ± 0.81) × $10^{-11}$ $cm^3$ $molecule^{-1}$ $s^{-1}$ and our extrapolated value for $k_{3a}$(10 K) = 3.5 × $10^{-11}$ $cm^3$ $molecule^{-1}$ $s^{-1}$. The large discrepancy between these two studies at the lowest temperatures merits some discussion given the potential importance of this reaction to dense cloud chemistry. The measurements of reaction (3a) by Brownsword *et al.*[15] were performed using the same cryogenically cooled slow flow apparatus as with an earlier study of reaction (1) and the O + OH reaction by Smith and Stewart.[42] In contrast to the current work employing the relative rate method, these experiments used reaction (2) to quantify the excess atom concentration directly through observation of the infrared chemiluminescent emission from the $N_2$ product to determine the titration end point. As it was not possible to perform these critical titration measurements in the detection region, they were performed instead at a distance of 30 cm downstream, outside of the cryogenically cooled part of the apparatus. The observed divergence between the results of these cryogenically cooled experiments and the more recent CRESU experiments of the present reaction, the N + OH reaction and the O + OH reaction could have originated from an overestimation of the atom wall losses at low temperature; a problem which would have been exacerbated by the distance between the detection region and the titration point. Although it is clearly not possible to confirm this hypothesis, the available evidence seems to weigh in favour of a systematic error in these earlier measurements. Indeed, whilst CRESU type experiments (which are unaffected by wall losses) of atom – unstable radical reactions (1), (3a), (4) and the O + OH reaction employing both relative[6,10] and absolute[43] methods indicate that the rate constants for these processes show a much less pronounced temperature dependence above 50 K, in good agreement with the calculations of several different groups employing a variety of classical and quantum mechanical methods,[6,11,44-47] the results of Brownsword *et al.*[15] and Smith and Stewart[42] predict negative temperature dependences, leading to large rate constant values at temperatures pertinent to dense clouds. Could part of the explanation for these differences lie with the present and earlier relative rate measurements which might conceivably be strongly affected by secondary reactions?

Whilst every effort has been made to reduce secondary reactions to a minimum in the present work, through the use of low precursor molecule concentrations, it is possible that such processes could lead to significant errors in the measured rate constants. Nevertheless, the fact that our measurements of the relative rate of the N + CH reaction with respect to both reactions (1) and (4) yield absolute rate constants which are in good agreement with each other and with the calculated ones suggests that this is not the case, allowing us to rule out secondary reactions as a major factor in the present experiments. Another potential explanation for the differences between these relative rate measurements and the earlier absolute ones by Smith and coworkers could be the value of the rate constant attributed to the N

+ NO reaction at low temperature as measured using the CRESU technique by an absolute method with the NO concentration given by its partial pressure.[7] As this process was used as the initial reference reaction to obtain temperature dependent rate constants for reaction (1), which itself has been used as the reference for subsequent work (including the present investigation),[6,10] significant errors in the N + NO rate constant would be incorporated into the rates of all subsequent target reactions. Let us test this hypothesis by assuming initially that the low temperature measurements by Smith and coworkers are correct. Smith and Stewart measured a rate constant for reaction (1) $k_1$(103 K) = (8.0 ± 0.8) × $10^{-11}$ $cm^3$ $molecule^{-1}$ $s^{-1}$. Using the relative rate value $k_1/k_2$ obtained by Daranlot *et al.*[6] of approximately 1.1 ± 0.1 at this temperature to evaluate $k_2$ gives a value for $k_2$(103 K) = 7.3 × $10^{-11}$ $cm^3$ $molecule^{-1}$ $s^{-1}$ which is almost a factor of two larger than the one given by the fitting parameters of Bergeat *et al.*[7] ($k_2$(103 K) = 4.1 × $10^{-11}$ $cm^3$ $molecule^{-1}$ $s^{-1}$) and outside of the range of possible values given by the experimental error bars. Nevertheless, supplementary measurements of the low temperature rate constants for reaction (2) would be extremely helpful to resolve this issue.

## 4. Conclusions and Astrophysical Implications

Rate constants for the N($^4$S) + CH($X^2\Pi_r$) reaction by both experimental and theoretical methods are reported here for a wide range of temperatures, extending down to 10 K. The reaction is shown to proceed through a complex forming mechanism involving the two lowest lying triplet ($1^3A'$ and $1^3A''$) states of HCN/HNC leading to CN formation in both the ground $X^2\Sigma^+$ and first excited $A^2\Pi$ states. The experimental and theoretical rate constants are in excellent agreement, predicting a positive temperature dependence for this reaction. Whilst the present results are in good agreement at room temperature with an earlier study using a cryogenic cooling method, the two datasets diverge at lower temperature leading to drastically different extrapolated values for the rate constant at temperatures pertinent to dense interstellar clouds. In order to test the impact of the new rate constant for reaction (3a) on CN production (and $N_2$ production indirectly) in dense interstellar clouds, we have used the Nautilus gas-grain model. This model computes the chemical abundances in the gas-phase and at the surface of the ices of interstellar grains as a function of time under the extreme conditions of the interstellar medium. Details on the processes included in the model can be found in Hersant *et al.*[48] and Semenov *et al.*[49] The kinetic data used for the simulations are the same as in those used in Daranlot *et al.*[10] All other parameters (physical conditions, elemental abundances and initial chemical composition) are the same as in Daranlot *et al.*[10] for typical dark molecular clouds. Two values of the oxygen elemental abundance have been considered (2.4 × $10^{-4}$ and 1.4 × $10^{-4}$ with respect to total hydrogen (nH + 2n$H_2$)) resulting in C/O elemental ratios of 0.7 and 1.2. Figure 6 shows the relative abundances of CN and CH as a function of time obtained using our earlier model[10] (using a rate constant $k_{3a}$(10 K) = 2.3 × $10^{-10}$ $cm^3$ $molecule^{-1}$ $s^{-1}$) compared to those obtained when we use the new value for $k_{3a}$(10 K) = 2.5 × $10^{-11}$ $cm^3$ $molecule^{-1}$ $s^{-1}$. Surprisingly, we observe only small changes in the relative abundances of both CH and CN using this substantially lower rate constant, leading to a decrease in the predicted $N_2$ abundance at $10^5$ years of between 10 and 20 %, suggesting that this reaction is in fact only a relatively minor source of CN radicals in dense clouds, in contrast to previous assumptions. A closer examination of the model output, however, highlights the reactions of nitrogen with several unsaturated long chain carbon radicals as important CN formation mechanisms. As the chemistry of these species is described by only a limited network of reactions which are assigned inconsistent low temperature rates, these sources of CN are likely to be heavily overestimated in the present model. Indeed, most current models generally overestimate the observed CN abundances in dense clouds. Work is underway to address these issues in order to provide a more reliable description of interstellar CN formation. Among other candidates as potential sources of CN are reactions (3b) and (3c) given the predicted high abundances of both of these species. Future work will address the low temperature reactivity of these two reactions to assess their relative importance as CN formation mechanisms using an updated dense cloud model.

**Acknowledgements**

The experimental work was supported by the Agence Nationale de la Recherche (ANR-JC08_311018), the Conseil Régional d'Aquitaine (20091102002) and the CNRS program EPOV. The experimental and modeling work were both supported by the INSU-CNRS national programs PCMI and PNP and the Observatoire Aquitain des Sciences de l'Univers. HG is supported by the Department of Energy (DE-FG02-98ER15694). DX was supported by the National Natural Science Foundation of China (Grant Nos. 21273104, 91221301, 91021010) and the Ministry of Science and Technology (Grant No. 2013CB834601).

[a]*Université de Bordeaux, Institut des Sciences Moléculaires, CNRS UMR 5255, F-33400 Talence, France. Tel: +33 5 40 00 63 42*

*E-mail: km.hickson@ism.u-bordeaux1.fr*

[b]*Institute of Theoretical and Computational Chemistry, Key Laboratory of Mesoscopic Chemistry, School of Chemistry and Chemical Engineering, Nanjing University, Nanjing 210093, China.*

*E-mail: dqxie@nju.edu.cn*

[c]*Université de Bordeaux, Laboratoire d'Astrophysique de Bordeaux, CNRS UMR 5804, F-33270 Floirac, France.*

[d]*Department of Chemistry and Chemical Biology, University of New Mexico, Albuquerque, NM 87131, USA.*

*E-mail: hguo@unm.edu*

## References

1 R. P. A. Bettens and M. A. Collins, *J. Chem. Phys*., 1998, **109**, 9728.

2 D. B. Milligan, D. A. Fairley, C. G. Freeman and M. J. McEwan, *Int. J. Mass Spectrom*., 2000, **202**, 351.

3 E. Herbst, D. J. DeFrees, and A. D. McLean, *Astrophys. J*., 1987, **321**, 898.

4 M. Womack, L. M. Ziurys and S. Wyckoff, *Astrophys. J*., 1992, **393**, 188.

5 D. C. Knauth, B.-G.Andersson, S.R. McCandliss and H. W. Moos, *Nature*, 2004, **429**, 636.

6 J. Daranlot, M. Jorfi, C. Xie, A. Bergeat, M. Costes, P. Caubet, D. Xie, H. Guo, P. Honvault and K. M. Hickson, *Science*, 2011, **334**, 1538.

7 A. Bergeat, K. M. Hickson, N. Daugey, P. Caubet and M. Costes, *Phys. Chem. Chem. Phys*., 2009, **11**, 8149.

8 M. Jorfi and P. Honvault, *J. Phys. Chem. A,* 2009, **113**, 10648.

9 P. Gamallo, R. Martínez, R. Sayós and M. González, *J. Chem. Phys.,* 2010, **132**, 144304.

10 J. Daranlot, U. Hincelin, A. Bergeat, M. Costes, J.-C. Loison, V. Wakelam and K. M. Hickson, *Proc. Natl. Acad. Sci. USA*, 2012, **109**, 10233.

11 J. Ma, H. Guo and R. Dawes, *Phys. Chem. Chem. Phys*., 2012, **14**, 12090.

12 V. Wakelam, E. Herbst, J.-C. Loison, I. W. M. Smith, V. Chandrasekaran, B. Pavone, N. G. Adams, M.-C. Bacchus-Montabonel, A. Bergeat, K. Béroff, V. M. Bierbaum, M. Chabot, A. Dalgarno, E. F. van Dishoeck, A. Faure, W. D. Geppert, D. Gerlich, D. Galli, E. Hébrard, F. Hersant, K. M. Hickson, P. Honvault, S. J. Klippenstein, S. Le Picard, G. Nyman, P. Pernot, S. Schlemmer, F. Selsis, I. R. Sims, D. Talbi, J. Tennyson, J. Troe, R. Wester, L. Wiesenfeld, *Astrophys. J. Suppl. Ser*., 2012, **199**, 21.

13 E. Herbst, *OSU 09-2008 database*, 2008, https://www.physics.ohio-state.edu/~eric/research.html

14 J. Woodall, M. Agúndez, A. J. Markwick-Kemper and T. J. Millar, *Astron. Astrophys.*, 2007, **466**, 1197.

15 R. A. Brownsword, S. D. Gatenby, L. B. Herbert, I. W. M. Smith, D. W. A. Stewart and A. C. Symonds, *J. Chem. Soc. Faraday Trans*., 1996, **92**, 723.

16 U. Hincelin, V. Wakelam, F. Hersant, S. Guilloteau, J.-C. Loison, P. Honvault and J. Troe, *Astron. Astrophys.*, 2011, **530**, A61.

17 P. Pratap, J. E. Dickens, R. L. Snell, M. P. Miralles, E. A. Bergin, W. M. Irvine and F. P. Schloerb, *Astrophys J.*, 1997, **486**, 862.

18 J. E. Dickens, W. M. Irvine, R. L. Snell, E. A. Bergin, F. P. Schloerb, P. Pratap and M. P. Miralles, *Astrophys J.*, 2000, **542**, 870.

19 B. R. Rowe, G. Dupeyrat, J. B. Marquette and P. Gaucherel, *J. Chem. Phys.*, 1984, **80**, 4915.

20 B. Vidal and C. Dupret, *J. Phys. E.: Sci. Instr.*, 1976, **9**, 998-1002.

21 N. Daugey, P. Caubet, B. Retail, M. Costes, A. Bergeat and G. Dorthe, *Phys. Chem. Chem. Phys.*, 2005, **7**, 2921-2927.

22 A. Canosa, I. R. Sims, D. Travers, I. W. M. Smith and B. R. Rowe, *Astron. Astrophys.*, 1997, **323**, 644.

23 M. A. Blitz, M. Pesa, M. J. Pilling and P. W. Seakins, *J. Phys. Chem. A* 1999, **103**, 5699.

24 S. Zabarnick, J. W. Fleming and M. C. Lin, *21st Symp. (Int.) on Combust.*, The Combustion Institute, Pittsburg, Pennsylvania, USA, 1986, pp. 713.

25 A. Bergeat, S. Moisan, R. Méreau and J.-C. Loison, *Chem. Phys. Lett.*, 2009, **480**, 21.

26 J. Daranlot, A. Bergeat, F. Caralp, P. Caubet, M. Costes, W. Forst, J.-C. Loison and K. M. Hickson, *ChemPhysChem,* 2010, **11**, 4002.

27 P. J. Knowles and H. J. Werner, *Chem. Phys. Lett.*, 1988, **145**, 514.

28 H. J. Werner and P. J. Knowles, *J. Chem. Phys.*, 1988, **89**, 5803.

29 S. R. Langhoff and E. R. Davidson, *Int. J. Quantum Chem.*, 1974, **8**, 61.

30 MOLPRO, version 2010.1, a package of ab initio programs, H.-J. Werner, P. J. Knowles, G. Knizia, F. R. Manby, M. Schütz, P. Celani, T. Korona, R. Lindh, A. Mitrushenkov, G. Rauhut, K. R.Shamasundar, T. B. Adler, R. D. Amos, A. Bernhardsson, A. Berning, D. L. Cooper, M. J. O. Deegan, A. J. Dobbyn, F. Eckert, E. Goll, C. Hampel, A. Hesselmann, G. Hetzer, T. Hrenar, G. Jansen, C. Köppl, Y. Liu, A. W. Lloyd, R. A. Mata, A. J. May, S. J. McNicholas, W. Meyer, M. E. Mura, A. Nicklass, D. P. O'Neill, P. Palmieri, K. Pflüger, R. Pitzer, M. Reiher, T. Shiozaki, H. Stoll, A. J. Stone, R. Tarroni, T. Thorsteinsson, M. Wang and A. Wolf, see http://www.molpro.net.

31 B. Fu, X. Xu and D. H. Zhang, *J. Chem. Phys.*, 2008, **129**, 011103.

32 H. Guo, *Int. Rev. Phys. Chem.*, 2012, **31**, 1.

33 S. Lin and H. Guo, *Phys. Rev. A*, 2006, **74**, 022703.

34 Z. Sun, X. Lin, S.-Y. Lee and D. H. Zhang, *J. Phys. Chem. A*, 2009, **113**, 4145.

35 J. Ma, S. Y. Lin, H. Guo, Z. Sun, D. H. Zhang and D. Xie, *J. Chem. Phys.*, 2010, **133**, 054302.

36 Y. Georgievskii and S. J. Klippenstein, *J. Chem. Phys.*, 2005, **122**, 194103.

37 S. D. Le Picard, A. Canosa, B. R. Rowe, R. A. Brownsword and I. W. M. Smith, *J. Chem. Soc., Faraday Trans.*, 1998, **94**, 2889.

38 M. T. Rayez, P. Halvick, J. C. Rayez, P. Millié and B. Lévy, *Chem. Phys.*, 1994, **188**, 161.

39 N. Daugey, Ph. D. Thesis, Université Bordeaux 1, 1995.

40 P. J. Knowles, H. -J. Werner, P. J. Hay and D. C. Cartwright, *J. Chem. Phys.* 1988, **89**, 7334.

41 I. Messing, S. V. Filseth, C. M. Sadowski and T. Carrington, *J. Chem. Phys.* 1981, **74**, 3874.

42 I. W. M. Smith and D. W. A. Stewart, *J. Chem. Soc. Faraday Trans.*, 1994, **90**, 3221.

43 D. Carty, A. Goddard, S. P. K. Köhler, I. R. Sims and I. W. M. Smith, *J. Phys. Chem. A*, 2006, **110**, 3101.

44 F. Lique, M. Jorfi, P. Honvault, P. Halvick, S. Y. Lin, H. Guo, D. Q. Xie, P. J. Dagdigian, J. Kłos and M. H. Alexander, *J. Chem. Phys.* 2009, **131**, 221104.

45 A. J. C. Varandas, *J. Chem. Phys.* 2013, **138**, 134117.

46 M. Jorfi, P. Honvault and P. Halvick, *Chem. Phys. Lett.*, 2009, **471**, 65.

47 D. Edvardsson, C. F. Williams and D. C. Clary, *Chem. Phys. Lett.*, 2006, **431**, 261.

48 F. Hersant, V. Wakelam, A. Dutrey, S. Guilloteau and E. Herbst, *Astron. Astrophys.*, 2010, **522**, A42.

49 D. Semenov, F. Hersant, V. Wakelam, A. Dutrey, E. Chapillon, S. Guilloteau, T. Henning, R. Launhardt, V. Piétu and K. Schreyer, *Astron. Astrophys.*, 2009, **493**, L49.

**Table 1** Measured rate constants for the N + CH reaction

| $T$ / K | $[N]_{max}$ / $10^{14}$ atom $cm^{-3}$* | Flow density / $10^{16}$ molecule $cm^{-3}$ | $N^{†}$ | $k_{3a}$ / $10^{-11}$ $cm^3$ $molecule^{-1}$ $s^{-1}$ | $k_{3a}'/k_1'$ $\equiv k_{3a}/k_1$ | $k_{3a}'/k_4'$ $\equiv k_{3a}/k_4$ |
|---|---|---|---|---|---|---|
| 56 ± 1 | 1.6 | 22.2 ± 0.3 (Ar) | 51 | 5.9 ± 0.8 | 1.32 ± 0.10 | |
| 86 ± 1 | 2.6 | 11.9 ± 0.1 (Ar) | 53 | 8.3 ± 1.0 | 1.84 ± 0.11 | |
| 146 ± 1 | 4.1 | 9.9 ± 0.1(Ar) | 76 | 10.4 ± 1.1 | 2.32 ± 0.08 | |
| 146 ± 1 | 4.5 | 9.9 ± 0.1 (Ar) | 29 | 9.1 ± 1.1 | | 1.39 ± 0.09 |
| 166 ± 2 | 3.3 | 9.4 ± 0.2 (Ar) | 55 | 10.9 ± 1.1 | 2.42 ± 0.08 | |
| 167 ± 1 | 2.0 | 8.5 ± 0.1 ($N_2$) | 59 | 10.8 ± 1.1 | 2.40 ± 0.06 | |
| 296 | 2.6 | 16.2 (Ar) | 57 | 13.8 ± 1.5 | 3.00 ± 0.09 | |
| 296 | 2.0 | 16.8 (Ar) | 29 | 12.1 ± 1.7 | | 1.37 ± 0.06 |

* Estimated from fits to the decay profiles.

† Number of individual measurements

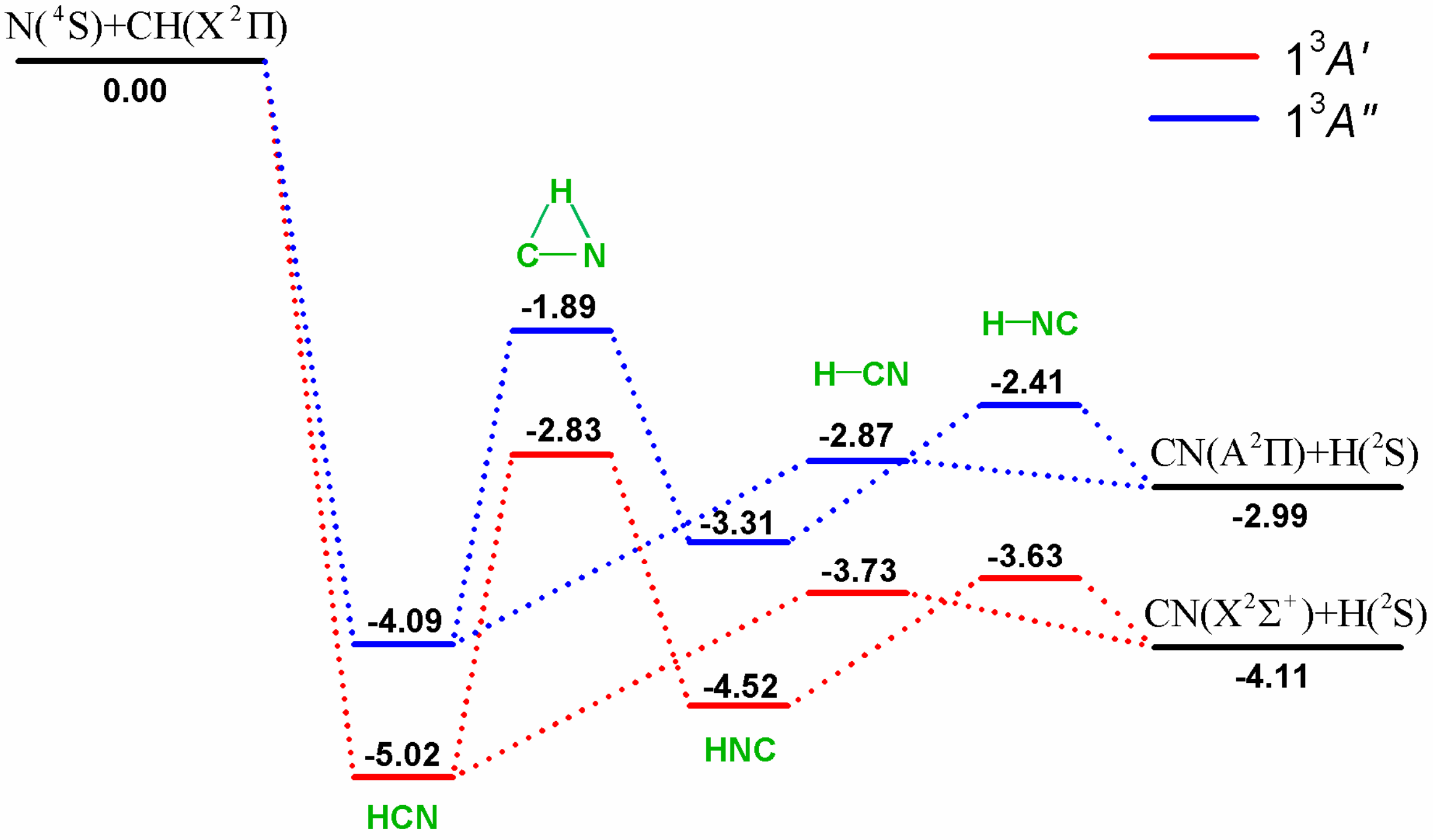

**Fig. 1**. Potential energy diagram for the reaction $N(^4S) + CH(X^2\Pi_r) \rightarrow H(^2S_g) + CN(X^2\Sigma^+, A^2\Pi_i)$. The calculated energies of the stationary points of HCN/HNC for the $1^3A'$ (red lines) and $1^3A''$ (blue lines) states are given in eV relative to the energy of the reactants.

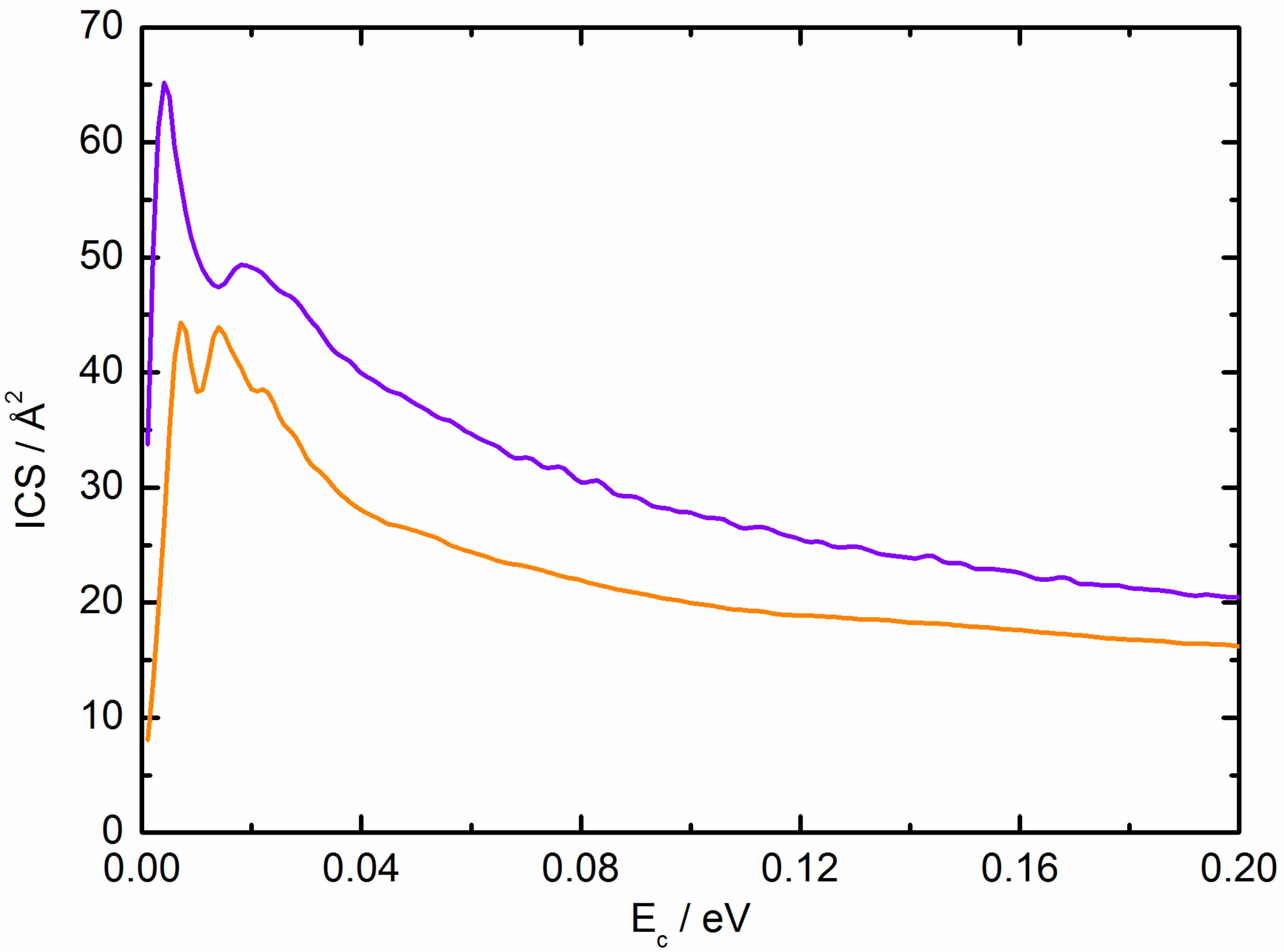

**Fig. 2**. Calculated integral cross sections (ICSs) for the N + CH reaction as a function of collision energy. (Purple solid line) calculated ICSs for reaction over the $1^3A''$ surface; (orange solid line) calculated ICSs for reaction over the $1^3A'$ surface.

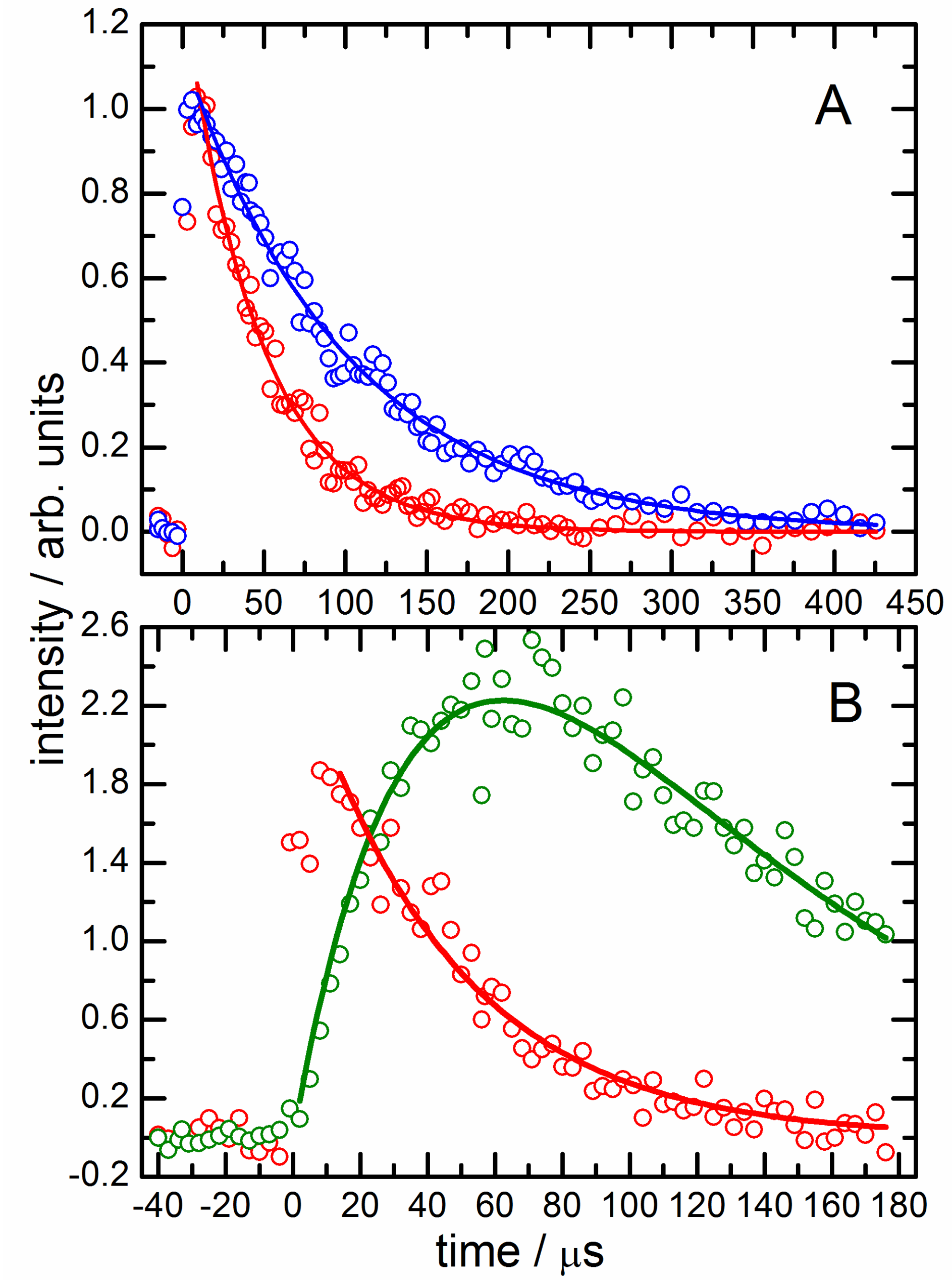

**Fig. 3**. (A) Variation of the fluorescence emission signal from reactant CH($^2\Pi_r$) and OH($^2\Pi_i$) radicals as a function of time recorded simultaneously at 296 K in the presence of an estimated excess of atomic nitrogen of 2.3 × $10^{14}$ atoms cm$^{-3}$ (B) As (A) for reactant CH($^2\Pi_r$) and product CN($^2\Sigma^+$) radicals recorded simultaneously at 146 K in the presence of an estimated excess of atomic nitrogen of 2.4 × $10^{14}$ atoms cm$^{-3}$.

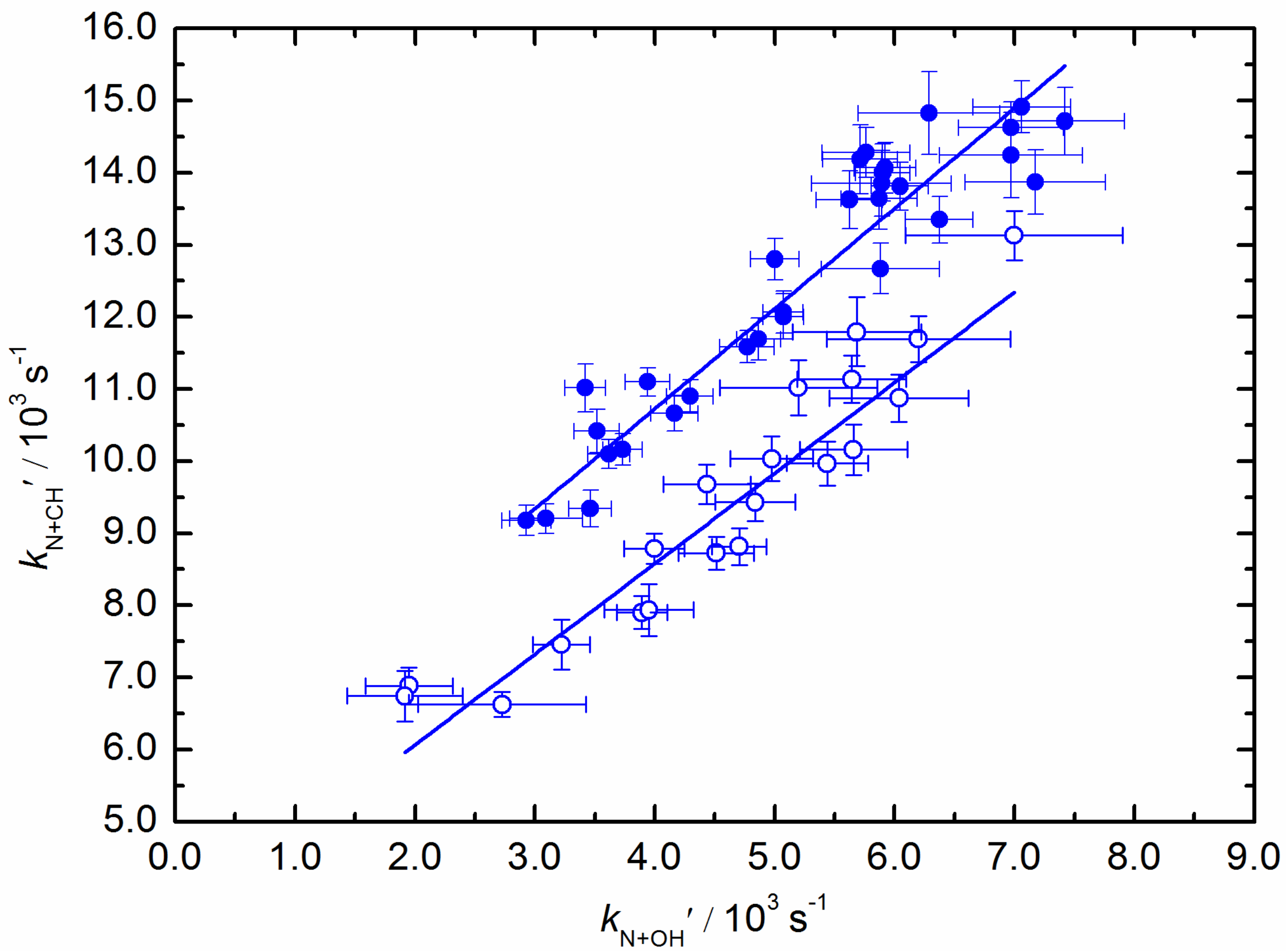

**Fig. 4**. Pseudo-first-order rate constants for reaction (3a) as a function of the pseudo-first-order rate constants for reaction (1) at 56 K. A weighted linear least squares fit yields the ratio of the second-order rate constants $k_{3a}/k_1$. The two plots have different y-axis intercept values due to the use of different precursor molecule concentrations to check for competing secondary reactions. The error bars on the ordinate reflect the statistical uncertainties at the level of a single standard deviation obtained by fitting to CH LIF profiles such as those shown in Fig. 3. The error bars on the abscissa were obtained in the same manner by fitting to the OH LIF profiles.

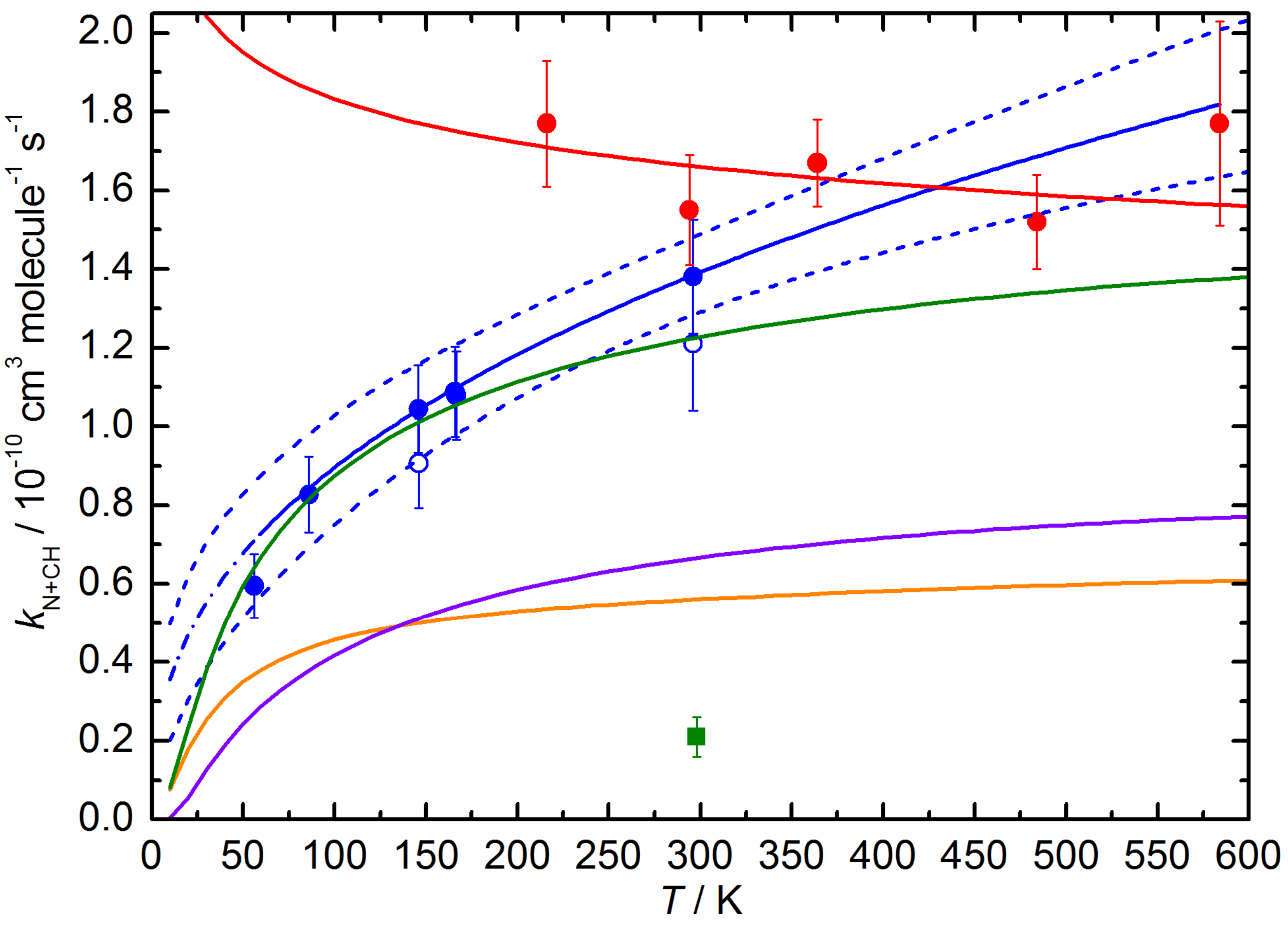

**Fig. 5**. Rate constants for the $N(^4S) + CH(^2\Pi_r)$ reaction as a function of temperature. (Red filled circles) Brownsword *et al.*[15]; (green filled square) Messing *et al.*[41]; (blue filled circles) this experimental work using reaction (1) as a reference; (blue open circles) this experimental work using reaction (4) as a reference; (blue dashed lines) 95 % confidence limits from fitting to the present experimental data and the data of Brownsword *et al.*[15] at room temperature and above; (solid blue line + dashed dotted line) fit to the current experimental data + extrapolation to 10 K; (purple solid line) this work, calculated rate constant for reaction over the $1^3A''$ surface; (orange solid line) this work, calculated rate constant for reaction over the $1^3A'$ surface; (green solid line) this work, calculated total rate constant; (red solid line) current recommendation from Wakelam *et al.*[12]

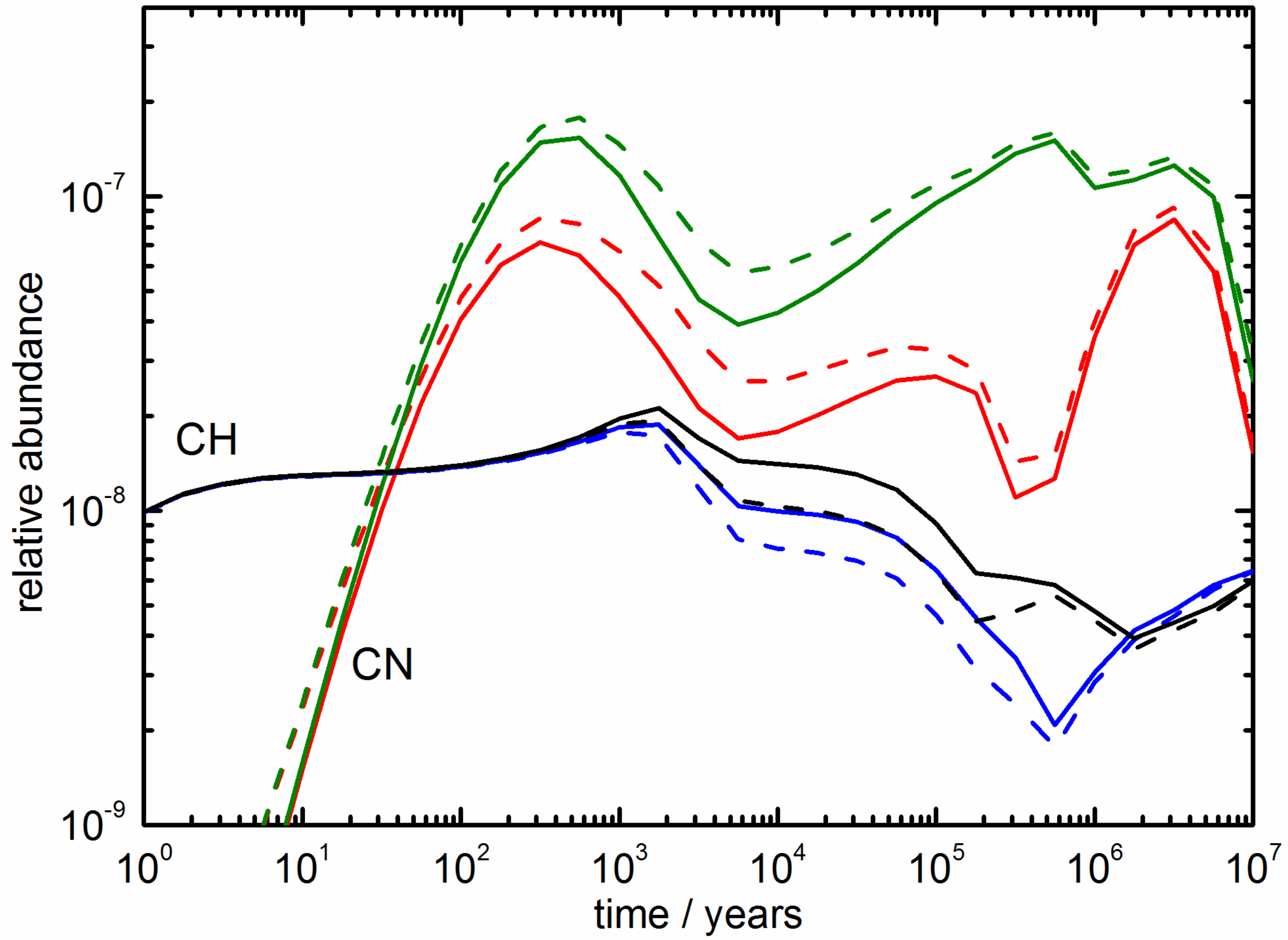

**Fig. 6**. Relative abundances of CN and CH as a function of time using a dense cloud model. (Dashed lines) simulations using a value for $k_{3a}$(10 K) = 2.3 × $10^{-10}$ $cm^3$ $molecule^{-1}$ $s^{-1}$; (solid lines) simulations using a value for $k_{3a}$(10 K) = 2.5 × $10^{-11}$ $cm^3$ $molecule^{-1}$ $s^{-1}$; (green lines) CN abundance with C/O = 1.2; (red lines) CN abundance with C/O = 0.7; (black lines) CH abundance with C/O = 1.2; (blue lines) CH abundance with C/O = 0.7.